\documentclass[preprint,showpacs,preprintnumbers,amsmath,amssymb]{revtex4}

\usepackage{graphicx}
\usepackage{dcolumn}
\usepackage{bm}

\begin{document}
\title{Stick slip motion in  grain grain friction in a humid atmosphere}

\author{O.Pozo$^1$, B.Soulestin$^2$,  N.Fraysse$^1$, N.Olivi-Tran$^2$}
\affiliation{$^1$ L.P.M.C., UMR-CNRS 6622, Universit\'e de Nice Sophia Antipolis,Parc Valrose, 06108 Nice cedex 2, France \\ $^2$S.P.C.T.S., UMR-CNRS 6638, E.N.S.C.I., 47 avenue Albert Thomas,
87065 Limoges cedex, France}

\date{\today}

\begin{abstract}
We set up an original apparatus to measure the grain grain friction
stress inside a granular medium composed of sodo-silicate-glass beads surrounded by a water vapor
atmosphere.We analyze here the influence of the physico chemistry of water
on our glass beads and its consequences on our shear experiment.
We found two scales in the analysis of the shear stress signal.
On the microscopic scale of one bead, the experimental results show
a dependence on the size of beads, on the shear rate and on humidity
for the resulting stick slip signal.
On the macroscopic
scale of the whole assembly of beads, the behavior of the total
amplitude of the shear stress depends on the size of the beads
and is humidity dependent only for relative humidity larger than 80\%.
For high degrees of humidity, on the microscopic scale, water
lubricates the surface of the beads leading to a decrease
in the microscopic resistance to shear while on the macroscopic
scale the resistance to shear is increased: the assembly of very
humid grains behaves as a non Newtonian fluid.

\end{abstract}

\maketitle

\pagebreak
\section{Introduction}
Granular materials are actually very simple: they consist of a great number
of macroscopic particles interacting only with their neighbours. In case of dry
samples, the sole interactions are hard core repulsion and friction.
However, in spite of their apparent simplicity, their behaviour can be very
complex and often surprising \cite{degennes,jaeger,jaeger2}.

Stick slip motion has been observed in various frictional experiments from solid-on-solid friction of rough surfaces \cite{heslot} to delicate studies
of lubricated films between smooth mica surfaces \cite{gee}.
Since one of the dominating interactions in granular materials
is friction among grains, it is not surprising that dense granular materials
exhibit various stick slip phenomena.
There are many examples of stick slip behaviour in granular materials.
In the first type of experiments stick slip sliding between a granular medium
and a solid body is considered \cite{duran,nasuno,nasuno2,geminard,ovarlez,lubert,deryck,frye,lacombe}.
In the second class of experiments the granular sample is forced
to deform slowly; an intermittent motion is caused by successive
discrete reorganization events inside the material. Experiments in elastic
annular shear cells revealed both regular \cite{cain} and irregular \cite{miller,dalton} stick slip behaviour.
In the third class, it is friction between grains themselves which is studied
\cite{coste}. The formation and breakage of chain forces are an explanation to the stick slip \cite{geng}. Moreover, comparison between a sheared granular
medium and avalanches in the same type of media has been made \cite{manna}.

Humidity changes the interactions between grains: for example
chemical reactions \cite{olivi,gayvallet} or mechanical behaviour
\cite{bocquet,fraysse,restagno}. 
Experiments have been made on the increase of the maximum angle of stability
of a humid granular media as a function of the quantity
of liquid added to the medium \cite{albert,barabasi} or as a function
of the humidity ratio of the surrounding atmosphere \cite{fraysse}.
In friction experiments, the effect of humidity has also been studied
\cite{ovarlez,frye,cain,dieterich}.

We will study here the effect of an under-saturated water vapor atmosphere
on the behaviour of a granular medium submitted to shear. The friction
studied here is a grain grain interaction obtained with an adapted apparatus.
The experimental set-up and experimental conditions will be detailed
in section II. Results will be discussed in section III. And finally
section IV contains the conclusion.

\section{Experimental Procedure}
\subsection{Set up}
We built an annular shear cell with an inner radius of 14 $mm$, an outer radius 
of 31 $mm$ and of 40 $mm$ depth. The shear cell is made of two parts:
one lower and one upper. The lower part remains motionless and has a depth
of 20 $mm$, while upper part rotates freely; this last has a depth of 20 $mm$
and is divided by four large vanes (see fig.1).
A gap of about 150 $\mu m$ is kept between the lower and the upper part
to prevent any friction between the two parts; this gap is small enough
to prevent the grains from escaping the shear cell. Granular matter
fills the lower part and half of the upper part. Such a shear apparatus
allows us to measure the shear force between the granular matter
contained in the lower part and the granular matter contained in the upper
part. This shear force is measured with a viscometer HAAKE VT550 (measuring
a 0.01-100 Pa range in shear forces for rotation speeds 0.1-800 rpm); moreover
the data sampling frequency is 5 to 10 times per second. A cover plate
introduces a small constant normal load (about 150 Pa) on the granular matter;
this cover is free to move vertically, which allows the dilatancy 
associated with the shear behavior.

The annular shear cell is kept inside a sealed heating unit, in which temperature (30$^oC \pm 1^oC$) is controlled by an adjustable water flow.
In order to control humidity, we used the method implemented
by Restagno et al. \cite{restagno} in which aqueous solutions saturated with
inorganic salts provide various levels of relative humidity (RH) in the atmosphere,
at a given temperature. For instance, to obtain a high humidity level
(95 \% RH), we used an ammonium dihydrogen phosphate ($NH_4H_2PO_4$)
solution and to obtain a low humidity level (about 5\% RH) we used silica gel.
 A small air pump forces the humid atmosphere to flow through the shear cell. We have checked
that the relative humidity is homogeneous throughout the shear cell. During
the experiments, relative humidity is measured inside and outside the shear cell using
two humidity sensors, having an accuracy of $\pm 1 \%$ RH.

\subsection{Preparation of the samples}

One type of granular materials has been used in the experiments: glass beads
(sodo silicate glass, Potters Ballotini).

Following is the procedure used for the measurements:

(1) The beads are washed with distilled water and a cleaning solution
 (RBS-25), then they are dried in an oven ( 12 hours at 130$^o$C). After that, the dried beads are sifted
using either two sieves of 315 and 400 $\mu$m pore size or
two sieves of 630 and 800 $\mu$m pore size. Note that the cleaning removes
the powder which is added to commercially available glass beads
in order to prevent the formation of clusters in presence of a humid atmosphere.
See for that fig. 3 which is a view taken with an optical
microscope using polarized light, of washed glass beads and non washed
glass beads: the non washed glass beads do not transmit polarized
light while washed beads do. This is the proof of the presence
of an additive on the surface of the glass beads.

(2) The beads are then poured into the sheared cell
and the cover plate is put. The granular matter is then
sheared in order to ensure the cover plate to be horizontal.

(3) The saline solution is then introduced in the experimental
set-up, and the set-up is kept aside for 7 hours in the temperature
controlled environment, to achieve a constant temperature and a homogeneous
distribution of humidity.

(4) Before starting the measurement, we rotate the shear cell for 40 s at an angular velocity
of 3.14 rad.s$^{-1}$. This defines the initial state of the granular sample.

(5) The shear stress is then measured for 600s at a small angular velocity.
Before each new measurement, the sample is changed.

\section{Results and Discussion}
\subsection{Results}
We studied the behaviour of the shear response of the granular medium,
at  several levels of humidity ranging from 0 to 100 \%. Shear stress is plotted
as a function of the dimensionless linear displacement $\delta d$, defined as $\delta d= (\frac{\bar{R} \omega}{\bar{d}_{bead}})t$, where $\omega$ is the angular velocity
of the viscometer, $\bar{R}$ the mean cell radius, $t$ the time
and $\bar{d}_{bead}$ the mean diameter of the beads.

We studied the system for 2 values of the angular velocity: 1.7.10$^{-3}$ rad/s
and 6.9.10$^{-3}$ rad/s.
At such small angular velocities,  the system is in a quasi static regime.

First of all, we have to make a difference between the mean amplitude MA of the stress
signal (difference between one local maximum and the very next local minimum of the signal) and the mean value MV (i.e. absolute value) of the same signal.

We observed a stick slip signal which corresponds
to the relaxation motion of the system (see fig.4).
Just after $t=0$ there is a transient regime where the stress increases
up to a maximum value and then decreases. 
This transient regime is not reproducible between two experiments:
the maximum of the shear stress depends on the initial configuration
of the beads.
Hence it is not possible to analyze the total cohesion of the system by the way of the maximum value
of the shear stress.

After this transient regime,
the permanent regime is reached (see fig. 4 again).
The stick slip signal is quite irregular due to the small individual
displacements of the beads.
 Note that there is an obvious difference between
the stick slip signal for a low humidity (13 \% RH) and for 
a high humidity (93 \% RH): the mean value MV of the stress signal for
a low humidity is  smaller than that for a high humidity and the amplitude MA
of the stress signal is larger for a higher value of the humidity.
Let us examine in details, in the following, these signals.

We computed the maximum and minimum values of each stick-slip event
during the permanent regime, i.e. after each stick event ( local increase
of the  signal), the local maximum
was saved and after each slip event ( local decrease of the  signal), the local minimum
 was saved. This allowed us to compute the averages of the local maxima
and local minima during the permanent regime over which we saved the stress
signal during 600$s$.
From these mean maximum and minimum, a mean amplitude MA was deduced
over the same duration of the permanent regime.

Fig.5 corresponds to the mean maximum and minimum as a function of
relative humidity RH. The diameter of the beads used here was
enclosed between 630 and 800 $\mu$m, while the angular velocity
of the shear cell was equal to 1.7.10$^{-3}$ rad/s.
The curve corresponding to minima is first quasi constant then
increases abruptly from a value of 88 \% RH on. The curve corresponding
to maxima increases regularly until the largest value of the stress
is reached.

Fig.6 corresponds also to the mean maximum and minimum as a function
of relative humidity RH, but this time, the diameter of the beads
was enclosed between 310 and 400 $\mu$m and the angular
velocity was equal to 6.9.10$^-3$ rad/s.
Here, on the contrary to fig.5, the two curves corresponding
respectively to maxima and minima follow the same behavior
and increase first with a small slope then from value of 85\% RH on, with a large slope.

To understand the differences between the preceding behaviors,
let us examine fig.7 and fig.8. Fig.7 corresponds to the mean amplitude MA
of the stick slip signal as a function of relative humidity RH
for 630-800$\mu$m diameter large beads.
Two curves have been traced on fig.7:  the mean amplitude MA
(for a angular velocity equal to 1.7.10$^{-3}$ rad/s) is first constant
then increases from RH=55\% on and finally decreases from RH=90 \% on.
The other curve (corresponding to a angular velocity equal
to 6.9.10$^{-3}$ rad/s) has a similar behaviour, with a maximum
which is lower.
Fig. 8 corresponds to the mean amplitude MA of the stick slip signal
as a function of relative humidity RH for 315-400 $\mu$m diameter large beads.
Here again the same velocities as for fig.7 have been imposed to the shear cell.
We observe a similar behavior of the two curves which have been traced on fig.8
than that on fig.7.
\subsection{Discussion of results in the light of the geometry of the assembly of beads}
The difference of amplitudes (MA) between two experiments comes from the number of contacts
inside the granulate: if this number is relatively large, the stick stage
will have a longer duration (because the granulate can resist
a larger shear stress) and hence, the observed stress will
be larger until the end of the stick stage where the granulate
begins to relax.

In the light of these preceding 2  figures (Fig. 7 and Fig. 8), we can say first,that there
is a velocity dependent behavior of our beads under shear, second,
that the size of the beads has also an influence on their behavior.
We can analyze the role of the angular velocity in terms of number of contacts
 between the beads. For the largest angular velocity that we used,
 the number of constant contacts (number of contacts which are not lost during one stick stage) between the beads
is smaller as the contacts have a larger probability to be lost during
the shear process due: i.e. the number of lost contacts per second is larger.
This explains why the maximum mean amplitude MA for RH=90\%, $\bar{d}_{bead}=630-800\mu$m
is lower for $\omega$=6.9.10$^{-3}$ rad/s than for $\omega$=1.7.10$^{-3}$ rad/s (Fig. 7).
Similarly, the size of the beads plays the same role in terms of number of constant contacts between the beads for one stick stage: if the glass beads are larger, the total number
of contacts between particles  is smaller inducing a smaller resistance to shear and hence
smaller amplitudes MA of the stick-slip stress (Fig. 5 and 6). These two behaviors have an
effect on a very short time corresponding to a microscopic length.

Let us examine now the behavior of the mean value  MV,
i.e. the action of shear on humid beads on the macroscopic level.
If we compare the mean values of the maxima of the MV (resp. minima)
in figures 5 and 6, the MV is larger for larger beads. So we can say that
the macroscopic resistance to shear is larger for larger beads.
Humidity has an effect on this macroscopic behavior
only for high degrees of humidities. We will analyze this last behavior
in section V-C.

\subsection{Physico chemistry of water on glass}
Let us recall some previous results about the interaction
between water and sodo silicate glass beads \cite{olivi,gayvallet,iler,ehrenstein}.
Under the influence of humidity or water  a reaction of the glass bead
surface with water follows that forms a 'water skin'. The alkali ions of the glass surface
are taken out of the silicic acid structure and are replaced by protons. So grows a permanent water-skin that cannot be taken away, not even under a vacuum. With
longer influence of water there grows, in addition, a very strongly adhering
temporary water-skin, which thickness depends on the temperature and the degree
of humidity, which can be taken away by drying. Through the presence of water,
the strength of the glass bead surface can be reduced.
Note also that our glass beads have been washed with a commercial cleaning solution
(RBS-25).  This cleaning solution is basic.
The attack of bases on glass increase with increasing temperature and time.
Our glass beads have been washed in a aqueous solution containing
90 \% distilled water and 10 \% cleaning solution (RBS-25) for $t=1h$
at ambient temperature.
The resulting pH of the cleaning solution was equal to 12.
Roughly, the following reactions occur under the action of alkali
at the glass surface:
\begin{equation}
-Si-O-R+H_2O \rightarrow -Si-OH+R^++OH^-
\end{equation}
where $R=Na,K, Li$ and others.
And:
\begin{equation}
-Si-O-Si-+OH^- \rightarrow -Si-OH+-Si-O-
\end{equation}
The rate of corrosion (resp. dissolution) is not determined
by a diffusion controlled process but by the active dissolution
of the $SiO_2$ network \cite{ehrenstein}. The loss of mass is proportional to time.
The smooth surface of the beads under the two water-skins (permanent and temporary) gets rough.
This is analogous to an aging process, but not during the shear experiment     itself, but during the preparation of the granular sample.
 
The surface of the glass
 beads was also attacked by the basic cleaning solution: the silicic structure
of the surface allows water to penetrate more profoundly
and hence leading to a thicker adsorbed water layer.

\subsection{Discussion: microscopic scale}
On the microscopic scale,  there are three regimes
occurring during the shearing experiment and as a function
of humidity.Here we will follow
the theory of Halsey and Levine \cite{halsey}.
We can say, that for RH lower than 50\%, the quantity of water
on the surface of the beads is not sufficient to change their behavior
during shear compared to the shearing of 'dry' beads.The
friction process between the beads is dominated by the deformation
of the nano asperities on the surface of the beads and the resistance
to shear of the beads surface is negligible \cite{halsey}.
Between RH=50\% and RH=90\%, is the friction regime \cite{halsey}:  we can say
that the water adsorbed layer on the surface of the beads induce a non negligible resistance to shear on the microscopic scale  inducing an increase
in the amplitude MA of the stick slip signal. 
And finally between RH=90\% and RH=100\%, the quantity of water
on the surface of the beads is sufficient to obtain a lubrication
of the surface of grains (i.e. always on the microscopic scale), and hence, the microscopic resistance
to shear decreases, and so does the amplitude MA. In that last case,
the quantity of water on the surface of the beads dissolves
the sodo-silicate glass.
This behavior of the shear stress as a function of the relative humidity
leads to the conclusion that the the quantity of water increases
first cohesion on the scale of one bead and then decreases it, for short time intervals. 
\subsection{Discussion: macroscopic scale}
On the macroscopic scale, and for humidity ranging from 0 to 80 \%, the resistance to shear of  humid beads
is similar to that of dry beads (increased resistance to shear
for large beads with respect to small beads, if comparison is made
between fig 5 and 6). This behavior is
the consequence of geometrical congestion for relatively
monodisperse spherical beads.

Always on the macroscopic scale, the action of water on the beads is only visible 
when its quantity is sufficient (RH larger than 80 \%) to attack and lubricate
the surface of the beads.  Then 
the assembly of beads has its resistance to shear  increased (value of the MV
 of the shear signal, Fig. 5 and 6) when the quantity of water
is high: a comparison may be made with dense colloidal dispersions under shear.

It may seem contradictory that for high degrees of humidity, on the microscopic
scale water lubricates the surface of the beads leading to an increased 
resistance to shear on the macroscopic scale. This may be  a consequence
of the non Newtonian characteristics of very humid grains.
Let us compare the very humid glass beads with  colloidal dispersions. Each colloidal particle
has a repulsive interaction with other colloidal particles (the attractive
part of the interaction can be neglected), but dense colloidal dispersions
observe a solid like behavior under shear even if there is no attractive
interactions between the particles \cite{botet}: when no shear is applied
the colloidal dispersion is fluid like becoming solid like as soon as
shear is applied on it. 
Indeed, the behavior of dense colloidal dispersions is non Newtonian and
observes a transition from fluid like to solid like as a function of
the concentration of the colloidal dispersions and as a function of shearing rate \cite{botet}.When the colloidal dispersion is in the solid like state,
its resistance to shear is increased compared to the state where it is fluid like.

Similarly, the humid glass grains are lubricated and the microscopic
resistance to shear decreases when humidity is larger than 90\%: the beads
have a non attractive contact force   like colloidal particles.
The explanation of the apparently contradictory behaviors on microscopic
and macroscopic scales may be that the very humid granulate is non Newtonian
leading to a higher macroscopic resistance to shear: the assembly
of grains observes a transition from a divided media to a solid like state
under shear.

So we join the conclusion of Tegzes \cite{tegzes}, that for high degrees
of humidity, the assembly of grains behaves as a non Newtonian fluid
(a viscoplastic fluid).

\section{Conclusion}

There are two scales which appear in grain grain friction
for a shearing experiment on glass beads in a water vapor atmosphere.

On the microscopic scale, the effect of humidity can be separated
into three regimes: the asperity regime where humidity has a negligible
role, the friction regime where the surface area of contacts between beads
increases the cohesion of the granular medium and a lubricated regime
where the quantity of water is sufficient reduce the viscosity 
of the surface of the beads.
Moreover, there is a size effect in the cohesion of the granulate:
for large beads, the number of contacts per unit volume between particles is smaller
tan for small beads
leading to a smaller total cohesion. The effect of shear rate is similar:
for higher shear rates, the number of lost contacts per unit time increases
leading also to a smaller total cohesion.

On the macroscopic scale, the behavior  of the resistance to shear
is similar to that of dry grains if RH is lower than 80\%.
For RH larger than 80\%, the assembly of beads behaves as
a non Newtonian fluid becoming solid like under shear.

\pagebreak
\begin{figure}
\includegraphics[width=12cm]{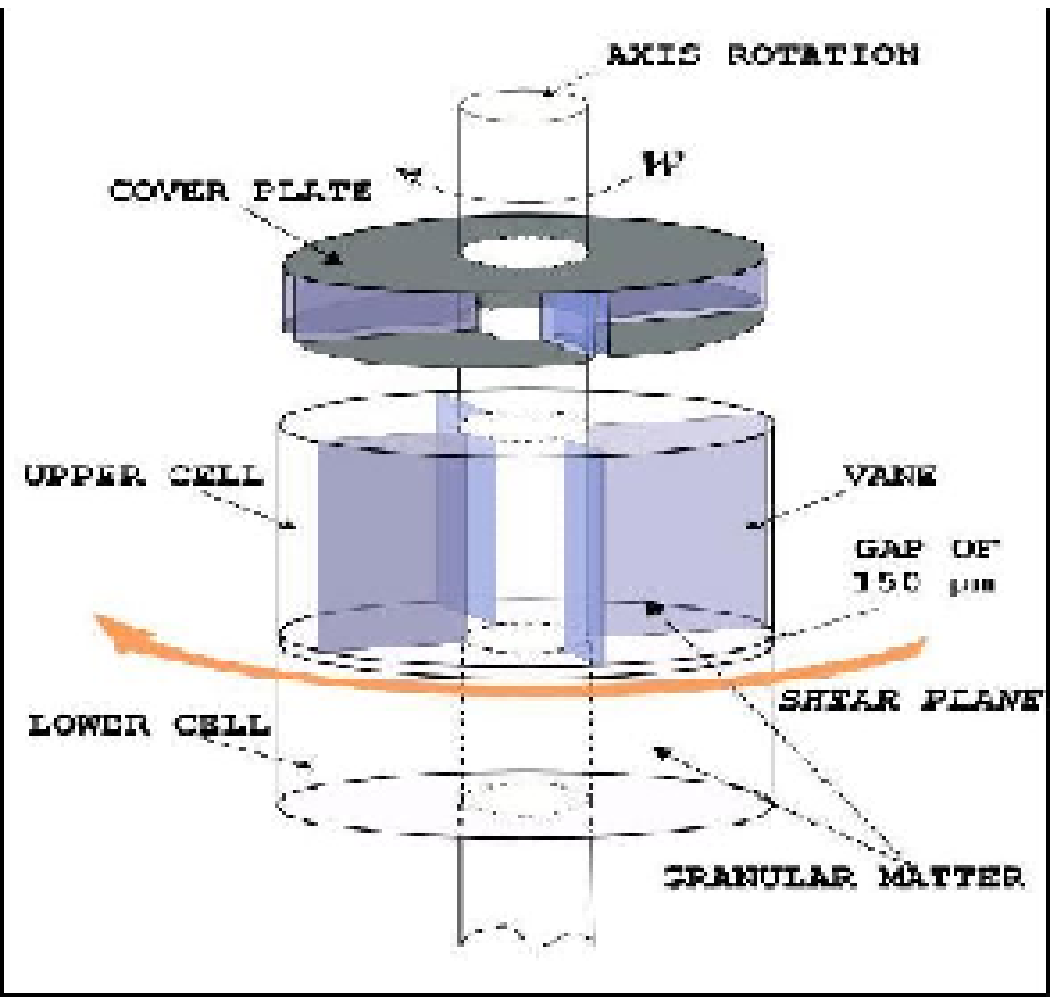}
\caption{Shear cell}
\end{figure}
\begin{figure}
\includegraphics[width=12cm]{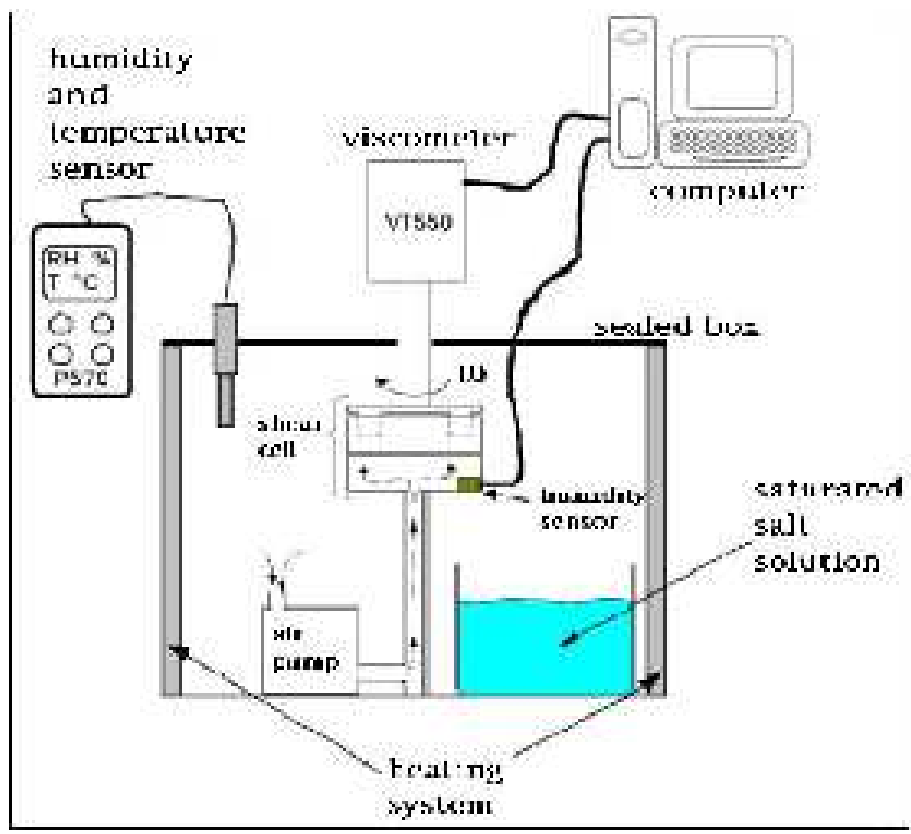}
\caption{Experimental set-up}
\end{figure}
\begin{figure}
\includegraphics[width=12cm]{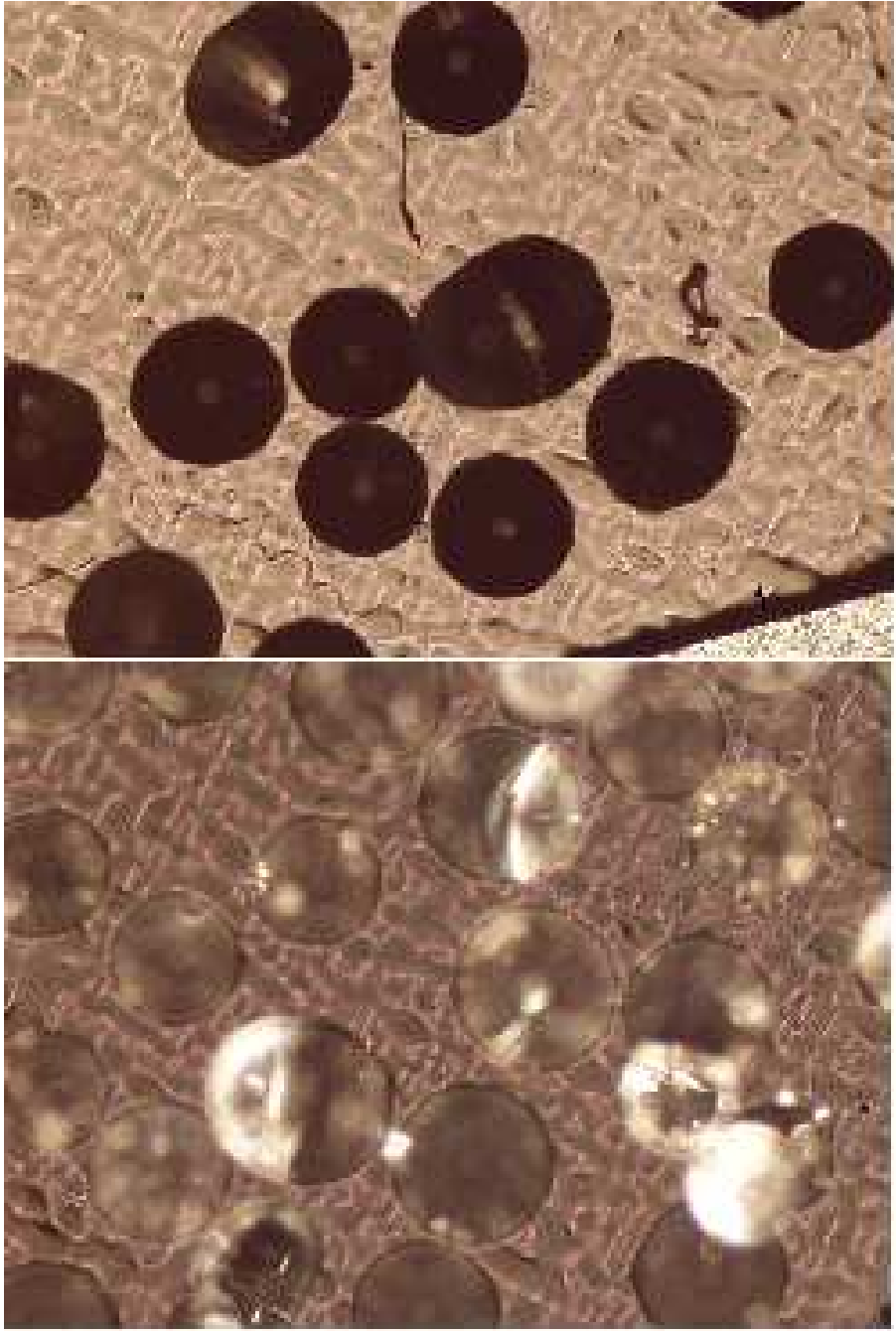}
\caption{
Top:Optical microscopy photography in polarized light of non washed commercial                                                                            sodo silicate glass beads
 Bottom: Optical microscopy photography in polarized light of  washed commercial
sodo silicate glass beads}
\end{figure}
\begin{figure}
\includegraphics[width=12cm,angle=270]{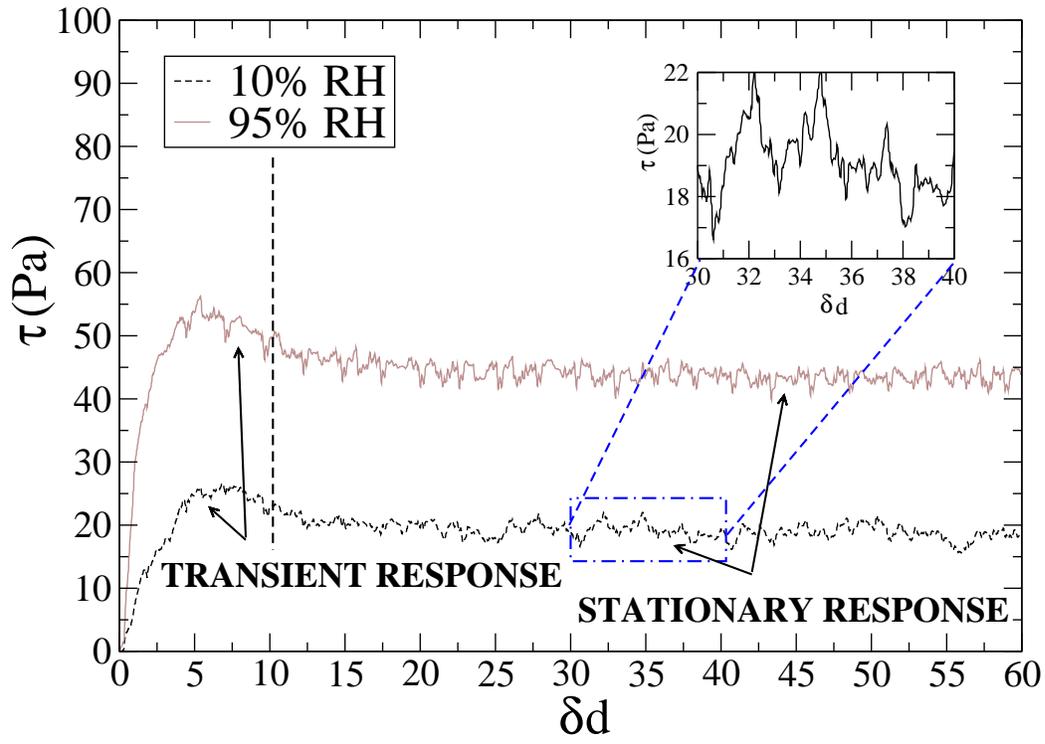}
\caption{Shear response as a function of $\delta d$ for a low and a high value
of humidity (13 \% and 93 \%). We can identify two regimes: a transient response
which is followed by a steady state.
}  
\end{figure}
\begin{figure}
\includegraphics[width=12cm,angle=270]{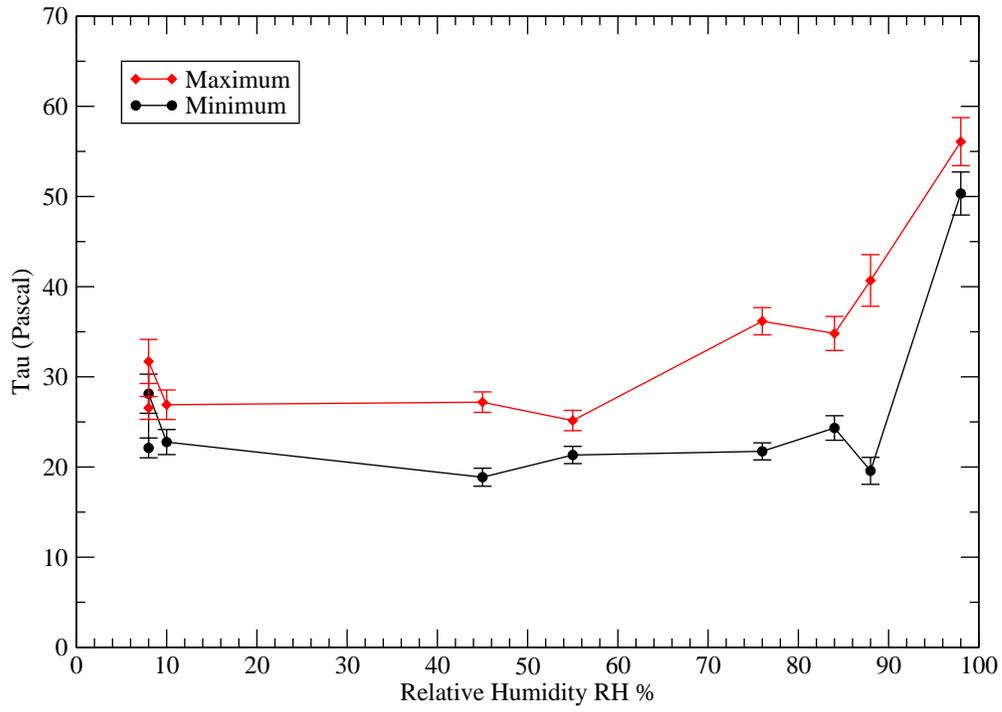}
\caption{Mean maxima and minima as a function of relative humidity. The maxima and minima
correspond to the steady state regime for an angular velocity $\omega=1.7.10^{-3}$ rad/s and a mean diameter of the beads equal to 630-800$\mu$m}
\end{figure}
\begin{figure}
\includegraphics[width=12cm,angle=270]{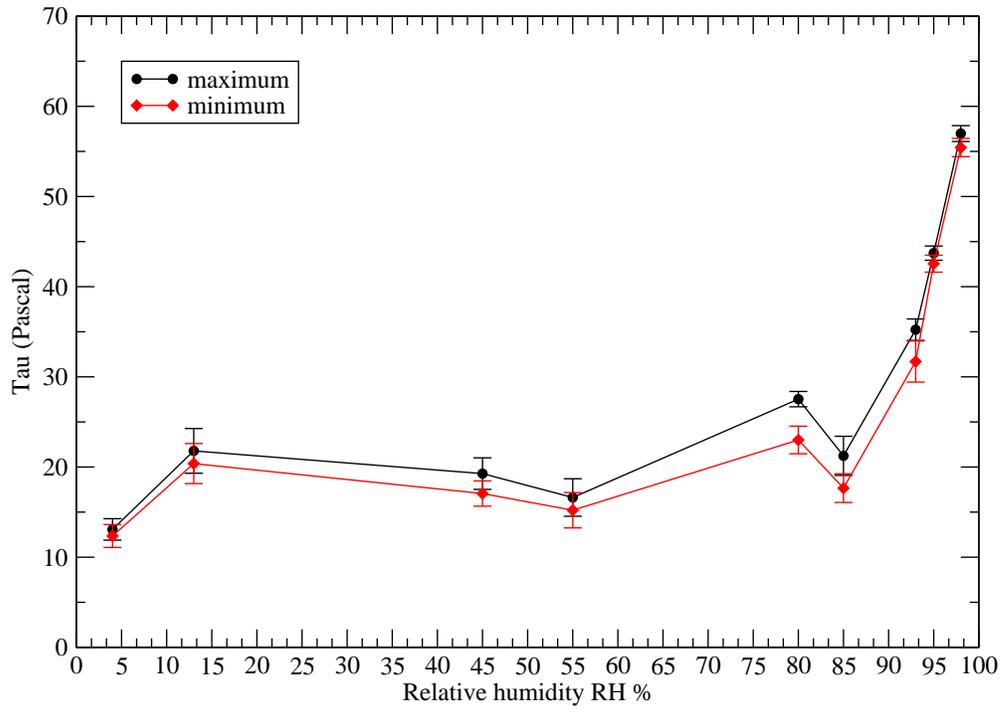}
\caption{Mean maxima and minima of the stress signal as a function of relative humidity RH. The maxima and minima
correspond to the steady state regime for an angular velocity $\omega=6.9.10^{-3
}$ rad/s and a mean diameter of the beads equal to 315-400$\mu$m}
\end{figure}
\begin{figure}
\includegraphics[width=12cm,angle=270]{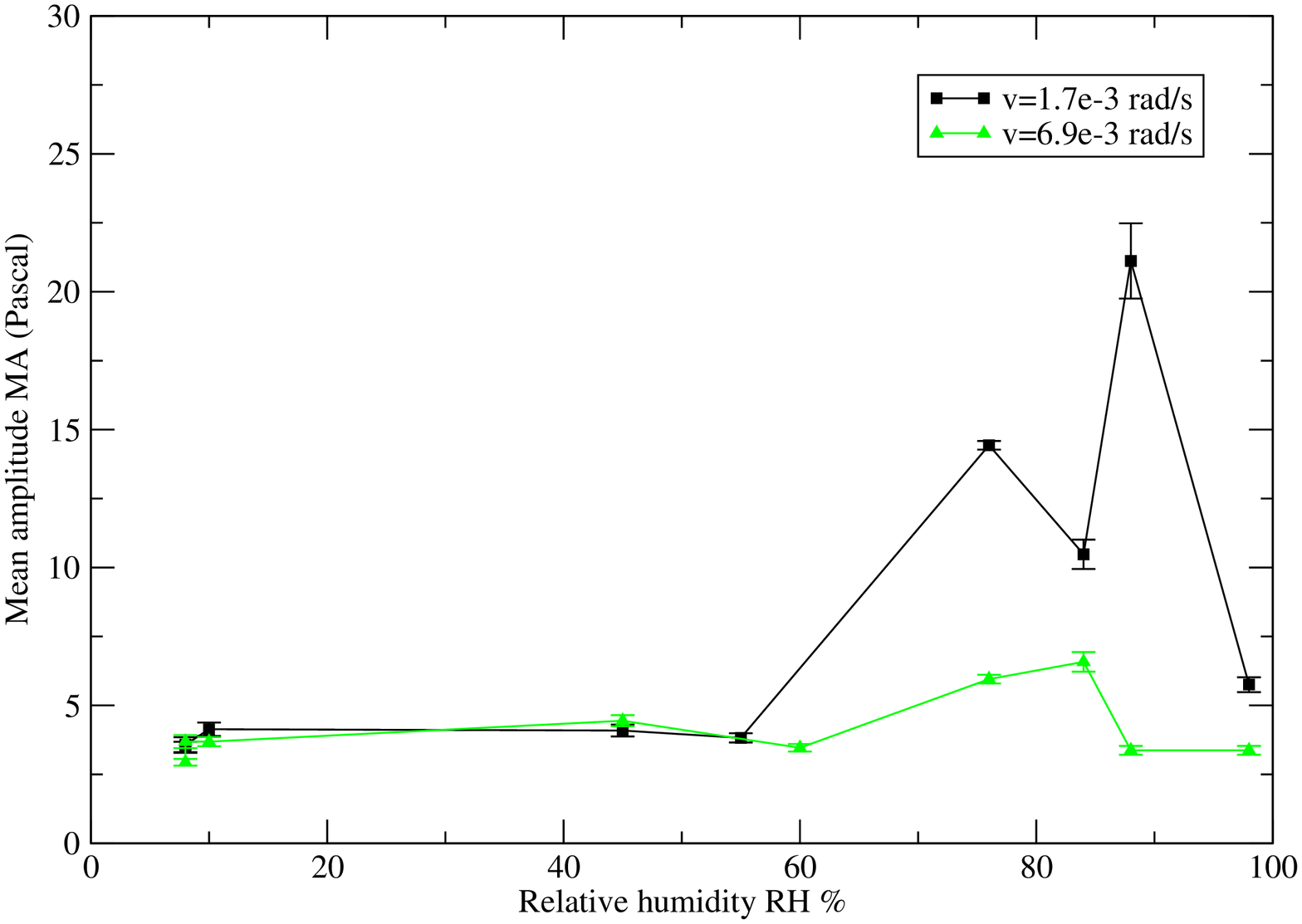}
\caption{Mean amplitude MA (difference between one local minimum and the very next maximum of the stress signal) of the stick slip signal corresponding to a mean diameter
of 630-800$\mu$m of the beads, as a function of relative humidity RH.
The squares correspond to a rotation velocity equal to $\omega=1.7.10^{-3
}$ rad/s, the triangles correspond to an angular velocity equal to $\omega=6.9.10^{-3}$ rad/s}
\end{figure}
\begin{figure}
\includegraphics[width=12cm,angle=270]{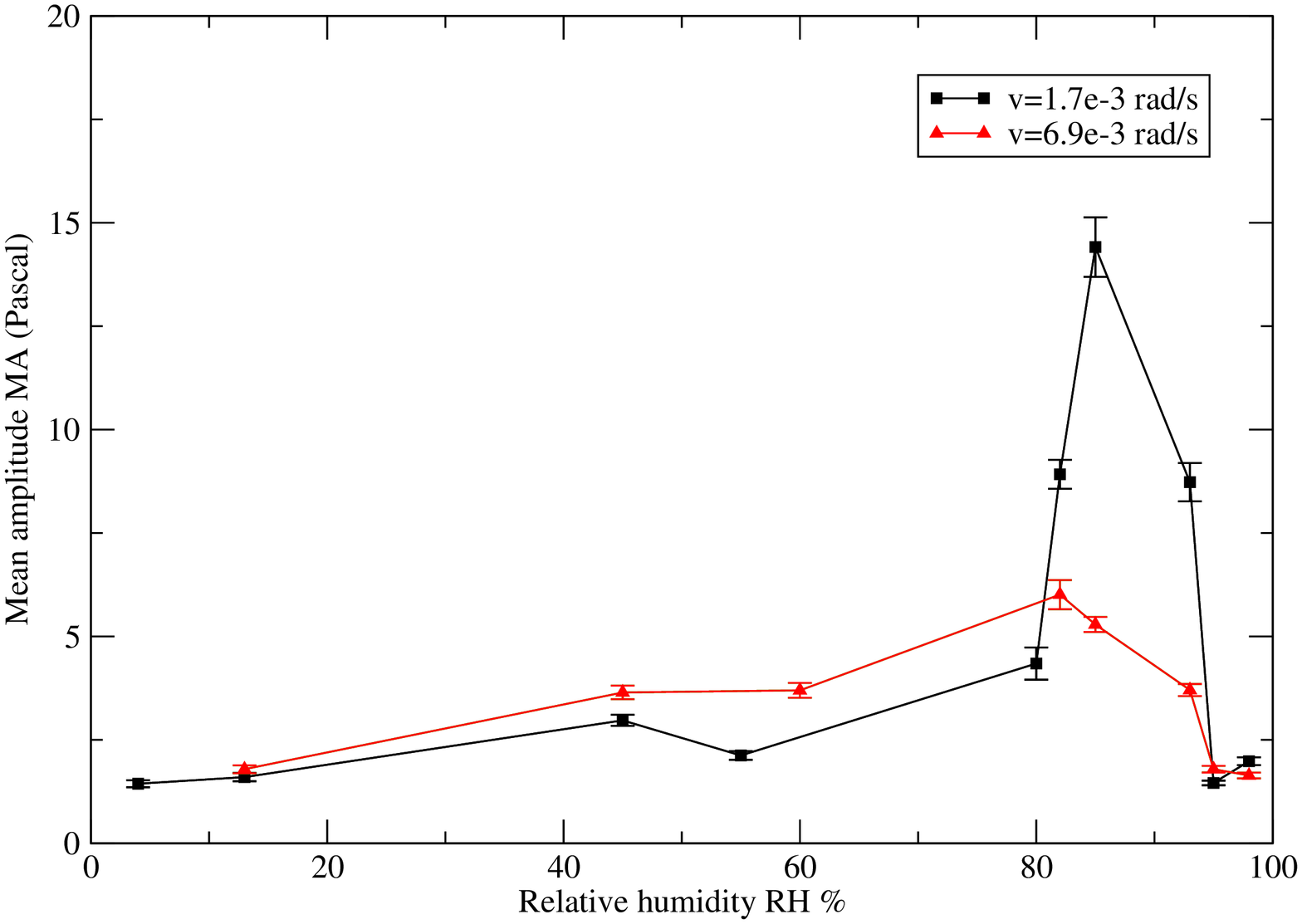}
\caption{Mean amplitude MA (difference between one local minimum and the very next maximum of the stress signal) of the stick slip signal corresponding to a mean diameter
of 315-400$\mu$m of the beads, as a function of relative humidity RH.The circles
correspond to a rotation velocity equal to $\omega=1.7.10^{-3
}$ rad/s, the diamonds correspond to an angular velocity equal to $\omega=6.9.10^{-3}$ rad/s}
\end{figure}

\end{document}